\documentclass[doublecol]{epl2}
\usepackage{amsmath,amssymb,graphics,epsfig,subfigure}

\title{Thermodynamic Geometry of black hole in the deformed Ho\v{r}ava-Lifshitz gravity}
\shorttitle{Thermodynamic Geometry of black hole in the deformed Ho\v{r}ava-Lifshitz gravity}

\author{Shao-Wen Wei\thanks{E-mail: \email{weishw@lzu.edu.cn}}
 \and Yu-Xiao Liu\thanks{Corresponding author. E-mail: \email{liuyx@lzu.edu.cn}}
 \and Yong-Qiang Wang\thanks{E-mail: \email{yqwang@lzu.edu.cn}}
 \and Heng Guo\thanks{E-mail: \email{guoh2009@lzu.edu.cn}} }
\shortauthor{S. W. Wei \etal}

\institute{
   Institute of Theoretical Physics,
    Lanzhou University,
    Lanzhou 730000, P. R. China
}

\abstract{We investigate the thermodynamic geometry and phase transition of {Kehagias-Sfetsos} black hole in the deformed Ho\v{r}ava-Lifshitz gravity with coupling constant $\lambda=1$. The phase transition in black hole thermodynamics is thought to be associated with the divergence of the capacities. {And} the structures of these divergent points are studied. We also find that the thermodynamic curvature produced by the Ruppeiner metric is positive definite for all $r_{+}>r_{-}$ and is divergence at $\eta_{2}=0$ corresponded to {the divergent points} of $C_{\Phi}$ and $C_{T}$. These {results} suggest that the microstructure of the black hole has an effective repulsive interaction, which is very similar to the ideal gas of fermions. These may shine some light on the microstructure of the black hole.}

\begin{document}

\maketitle

\section{Introduction}
\label{secIntroduction}

Motivated by Lifshitz theory in solid state physics \cite{Lifshitz1941},
Ho\v{r}ava proposed a new gravity theory at a Lifshitz point
\cite{Horava2009prd,Horava2009jhep,Horava2009}, referred as the
Ho\v{r}ava-Lifshitz (HL) theory. It has manifest three dimensional spatial
general covariance and time reparametrization invariance. This is a
non-relativistic renormalizable theory of gravity and it recovers the four dimensional general covariance only in an infrared limit. HL gravity provides an interesting classical and quantum field theory framework, where one can address some interesting questions and explore several connections to ordinary gravity or string theory.

The black hole solutions in the gravity theory have attracted much attention.
The spherically symmetric black hole solution with a dynamical parameter
$\lambda$ in asymptotically Lifshitz spacetimes was first given by L$\ddot{u}$, Mei and Pope \cite{Lu2009prl}. Subsequently, other black hole solutions and cosmological solutions were obtained and studied
\cite{cai2009prd,Kehagias2009plb}. The studies also focused on the thermodynamic properties and dynamical properties of different black hole
solutions \cite{Cao2009plb,bChen2009plb,Majhi2009,Peng2009,Myungf2009,wei2010,
aMyung2009}.

On the other hand, the Ruppeiner geometry \cite{Ruppeiner1979pra} is found to be a useful tool to study a thermodynamic system. It is generally considered to have physical meanings in the fluctuation theory of thermodynamics, and the components of the inverse Ruppeiner metric give second moments of fluctuations. The Ruppeiner geometry has been used to study the ideal gas and the van der Waals gas. The results show that the curvature vanishes for the ideal gas. While for the van der Waals gas, it is non-zero and divergent, {at which} the phase transitions take place \cite{Ruppeiner1995rmp,Ruppeiner00}. Thus the Ruppeiner geometry as a way to explore the thermodynamics and phase transition structure of black holes has been widely used \cite{Cai1999prd,Aman2003grg,Shen2007ijmpa,Aman2006grg,Mirzajhep2007,Medved2008mpla,
Quevedo2009prd,Sarkar2006jhep,CaoChen}. As pointed out in \cite{Ruppeiner00}, the Ruppeiner curvature can also be used to probe the microstructure of a thermodynamic system.

The purpose of this paper is to study the phase transition of the {Kehagias-Sfetsos (KS)} black hole \cite{Kehagias2009plb} in the HL gravity from the view of the thermodynamic geometry. The divergent points of the four capacities $C_{P}$, $C_{\Phi}$, $C_{T}$ and $C_{S}$ imply phase transitions in different ensembles. In order to understand the phase transition from the view of the thermodynamic geometry, we calculate the thermodynamic curvature. The result shows that {the divergent points} of the Ruppeiner curvature {are corresponded} to the divergence of capacities $C_{\Phi}$ and $C_{T}$. Thus, the information of phase transitions  {of the KS black hole} are contained in the thermodynamic {curvature}.

\section{Thermodynamics and phase transition of the Kehagias-Sfetsos black hole}
\label{ksblackhole}

The {KS} black hole in the deformed HL gravity reads \cite{Kehagias2009plb}
\begin{eqnarray}
 ds_{HL}^2 = - N^2(r)\,dt^2 + \frac{1}{f(r)}dr^2 + r^2 (d\theta^2
             +\sin^2\theta d\phi^2),
\end{eqnarray}
where
\begin{eqnarray}
 N^2=f=1 + \omega r^2-\sqrt{r(\omega^2 r^3+4\omega M)}.
 \label{solution}
\end{eqnarray}
It was argued in \cite{Peng2009,Myungf2009} that the quantity
$\sqrt{\frac{1}{2\omega}}$ behaviors as a charge-like parameter. So, we denote
$P=\sqrt{\frac{1}{2\omega}}$ and consider it as a new parameter in the black
hole thermodynamics. Then the metric function (\ref{solution}) will be of the
form
\begin{eqnarray}
 N^2(r,P)=f(r,P)=1+\frac{r^{2}}{2P^{2}}-\sqrt{\frac{r^{4}}{4P^{4}}+\frac{2Mr}{P^{2}}}.
\end{eqnarray}
Expanding the metric function at large $r$, {the Schwarzschild case will be recovered}. The outer (inner) horizon {of} the KS black hole is determined by $f(r,P)=0$, which gives
\begin{eqnarray}
 r_{\pm}=M\pm\sqrt{M^{2}-P^{2}}.\label{horizon}
\end{eqnarray}
Assuming the existence of the
black hole horizon ($M^{2}\geq P^{2}$), the mass parameter $M(r_{+},P)$ can be expressed
as
\begin{eqnarray}
 M=\frac{r_{+}}{2}+\frac{P^{2}}{2r_{+}}=\frac{r_{+}+r_{-}}{2}. \label{mass}
\end{eqnarray}
Note that the charge-like parameter $P$ satisfies $P^{2}=r_{+}r_{-}$. The Hawking temperature $T$ is defined as
\begin{eqnarray}
 T\!=\!\frac{f'(r)}{4\pi}\bigg|_{r=r_{+}}\!\!\!
   =\!\frac{r_{+}^{2}-P^{2}}{4\pi r_{+}(2P^{2}\!+\!r_{+}^{2})}
   \!=\!\frac{r_{+}-r_{-}}{4\pi r_{+}(r_{+}\!+\!2r_{-})}.\label{temperature}
\end{eqnarray}
At the extremal case $r_{+}=r_{-}$, the temperature $T$ vanishes. Assuming the first law of black hole thermodynamics $dM=TdS+\Phi dP$ holds, we then have the potential $\Phi$ corresponded to $P$
\begin{eqnarray}
 \Phi=\bigg(\frac{\partial M}{\partial P}\bigg)_{S}
             =\frac{r_{+}+2r_{-}-(r_{+}-r_{-})(\ln r_{+}^{2})}{r_{+}+2r_{-}}
                \sqrt{\frac{r_{-}}{r_{+}}}. \label{Phi}
\end{eqnarray}
From the first law, we can obtain the entropy of KS black hole:
\begin{eqnarray}
 S=\pi r_{+}^{2}+2\pi P^{2} (\ln r_{+}^{2})+S_{0},\label{1entropy}
\end{eqnarray}
where $S_{0}$ is an integration constant and can be fixed by the boundary
condition. The entropy can also be written in the form
\begin{eqnarray}
 S=\frac{A}{4}+2\pi P^{2} \ln \frac{A}{A_{0}}.\label{entropyarea}
\end{eqnarray}
Here $A=4\pi r_{+}^{2}$ is the outer horizon area and $A_{0}$ is a constant
with dimension of area. From (\ref{entropyarea}), one can see that the
Bekenstein-Hawking entropy/area law is modified by the second term in the
deformed HL gravity. As the charge-like
parameter $P\rightarrow 0$, {the standard Bekenstein-Hawking entropy/area law will be recovered}. The specific heat capacities for fixed charge $P$, $\Phi$ and the capacitances for fixed temperature $T$ and entropy $S$, in terms of $r_{+}$ and $r_{-}$, are given by
\begin{eqnarray}
 C_{P}\!\!\!&=&\!\!\!T\left(\frac{\partial S}{\partial T}\right)_{P}
             =\frac{2\pi r_{+}(r_{+}+2r_{-})^{2}(r_{+}-r_{-})}
              {\eta_{1}},\\
 C_{\Phi}\!\!\!&=&\!\!\!T\bigg(\frac{\partial S}{\partial T}\bigg)_{\Phi} \nonumber \\
         \!\!\!&=&\!\!\!-\frac{2\pi r_{+}}{\eta_{2}}\bigg[(r_{+}+2r_{-})^{3}
               +2\eta_{1}r_{-}(\ln r_{+}^{2})^{2}\nonumber\\
          &&\!\!\!+(r_{+}+2r_{-})(11r_{+}r_{-}-r_{+}^{2}+2r_{-}^{2})(\ln r_{+}^{2})\bigg],\nonumber\\ \\
 C_{T}\!\!\!&=&\!\!\!\bigg(\frac{\partial P}{\partial \Phi}\bigg)_{T}
         =-\frac{r_{+}(4r_{-}^{3}+12r_{+}r_{-}^{2}+3r_{+}^{2}r_{-}-r_{+}^{3})}
           {\eta_{2}(r_{+}-r_{-})},\nonumber\\ \\
 C_{S}\!\!\!&=&\!\!\!\bigg(\frac{\partial P}{\partial \Phi}\bigg)_{S}
        =\frac{r_{+}(r_{+}+2r_{-})^{3}}{\eta_{3}},
\end{eqnarray}
where
\begin{eqnarray}
 \eta_{1}\!\!\!\!&=&\!\!\!\! 5r_{+}r_{-}-r_{+}^{2}+2r_{-}^{2},\\
 \eta_{2}\!\!\!\!&=&\!\!\!\! 2r_{-}^{2}(2+(\ln r_{+}^{2}))
          +r_{+}r_{-}(16+5(\ln r_{+}^{2}))\nonumber\\
         &&\!\!\!\!+r_{+}^{2}(1-(\ln r_{+}^{2})),\\
 \eta_{3}\!\!\!\!&=&\!\!\!\!r_{+}^{2}r_{-}(9-2(\ln r_{+}^{2}))-r_{+}^{3})+r_{+}r_{-}^{2}(4r_{-}^{3}
   (1+(\ln r_{+}^{2})) \nonumber \\
        &&\!\!\!\!+(r_{+}+2r_{-})^{3}+2r_{+}r_{-}^{2}(12+5(\ln r_{+}^{2}))
            .
\end{eqnarray}
It is clear that the heat capacity $C_{P}$ approaches to zero as the black hole trends to the extremal case. It has a divergent behavior at $\eta_{1}=0$. Assuming a black hole is surrounded by the thermal radiation with the same temperature, the heat balance conditions will require the heat capacity of the black hole to be positive. This means that the positive heat capacity can guarantee a stable black hole to exist in thermal bath. While the negative one will make the black hole {disappear} when perturbation is included in.

\begin{figure*}
\centerline{\subfigure[]{\label{Cp}
\includegraphics[width=8cm]{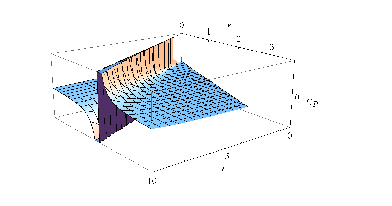}}
\subfigure[]{\label{Cphi}
\includegraphics[width=8cm]{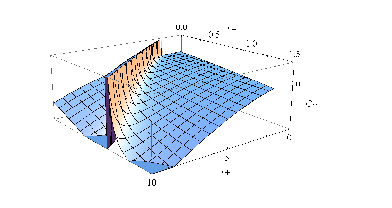}}}
\begin{center}
\subfigure[]{\label{CT}
\includegraphics[width=8cm]{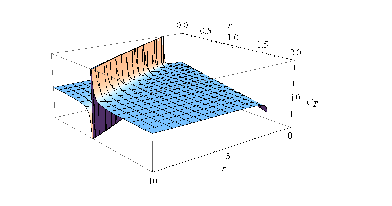}}
\subfigure[]{\label{CSA}
\includegraphics[width=8cm]{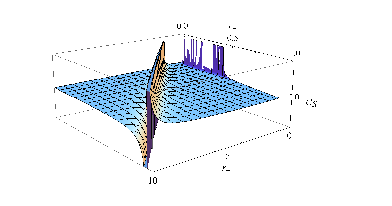}}
\subfigure[]{\label{CSB}
\includegraphics[width=8cm]{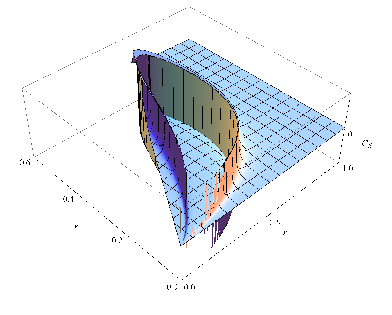}}
\end{center}
\caption{Behaviors of heat capacities for KS black hole in the range of $r_{+}\geq r_{-}$, (a) for $C_{P}$ with divergent points at $\eta_{1}=0$ and vanishing points at $r_{+}=r_{-}$, (b) for $C_{\Phi}$ with divergent points at $\eta_{2}=0$, (c) for $C_{T}$ with divergent points at $r_{+}=r_{-}$ and $\eta_{2}=0$, (d) for $C_{S}$ with divergent points at $\eta_{3}=0$, and (e) for $C_{S}$ at small $r_{+}$ and $r_{-}$.}\label{heatcapacity}
\end{figure*}

\begin{figure}
\begin{center}
\includegraphics[width=8cm]{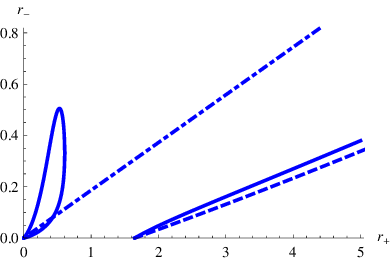}
\end{center}
\caption{Behaviors of phase transition points $\eta_{1}=0$ (dashed-dotted line), $\eta_{2}=0$ (dashed line) and $\eta_{3}=0$ (full line).}\label{eta}
\end{figure}

The behaviors of these four heat capacities are shown in Fig. \ref{heatcapacity} in the range of $r_{+}>r_{-}$. It is clear that each of them has divergent points. Form Fig. \ref{Cp}, we find that the small KS black holes are stable, while the large ones cannot exist stably in thermal bath. So, there exists a phase transition at $\eta_{1}=0$, where the heat capacity $C_{P}$ is divergent and changes its sign from positive to negative. It also clear that the capacities $C_{\Phi}$ and $C_{T}$ are divergent at $\eta_{2}=0$. $C_{T}$ is also found to go to negative infinity at $r_{+}=r_{-}$, which describes a phase transition from an extremal black hole to a non-extremal one. From the figures, we find that the capacity $C_{S}$ is divergent in two separate parts. One is similar to other capacities, and another part is a circle located at small $r_{+}$ and $r_{-}$ shown in Fig. \ref{CSB}. The different behaviors of them imply the thermodynamic stability and phase transitions in different thermodynamic ensembles. Form the definitions of the four capacities, we obtain a relation between them
\begin{eqnarray}
 C_{P}C_{T}(C_{\Phi}C_{S})^{-1}=1.\label{relation}
\end{eqnarray}
The relation tells us that only three of them are independent. To better understand the divergent behaviors of these heat capacities, we plot the phase transition points $\eta_{1}=0$, $\eta_{2}=0$ and $\eta_{3}=0$ in the $r_{+}\sim r_{-}$ plane presented in Fig. \ref{eta}.  We could see that the points $\eta_{1}=0$ and $\eta_{2}=0$ have a monotonically increasing behaviors, while points $\eta_{3}=0$ have a rich structure, i.e., a increasing line and a circle.

It is natural to conclude that the information of phase transition is contained in these capacities. And we will also show, in the next section, that the phase transition can also be revealed by the thermodynamic curvature produced by thermodynamic metric.

\section{Thermodynamic geometry of the Kehagias-Sfetsos black hole}
\label{ruppeiner}

The Ruppeiner metric, as we know, is defined as the second derivatives of entropy $S$ (thermodynamic potential) with respect to the mass and other
extensive quantities of a thermodynamic system. Different {from} the
Weinhold one, the Ruppeiner geometry is generally considered to have
physical meanings in the fluctuation theory of thermodynamics and
the components of the inverse Ruppeiner metric give second moments
of fluctuations. The Ruppeiner geometry as a way to explore the
black hole thermodynamics and phase transitions has been widely
used. Accordingly, how does the Ruppeiner geometry behave for
the KS black hole? Therefore, we start this section with the
question.

\begin{figure}
\centerline{
\includegraphics[width=8cm]{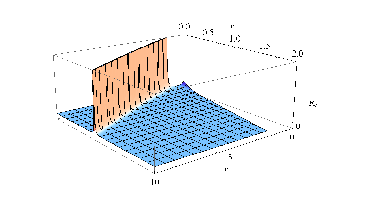}}
\caption{Behaviors of Ruppeiner curvature $R_{r}$ for KS black hole in the range of $r_{+}\geq r_{-}$, {with divergent points at $\eta_{2}=0$}.}\label{curvatures}
\end{figure}

The Ruppeiner thermodynamic metric for the KS black hole reads
\begin{eqnarray}
 ds_{r}^{2}=g_{ij}dx_{i}dx_{j}
           =-\frac{\partial^{2}S}{\partial x_{i}\partial x_{j}}
             dx^i dx^j,\;\;(i,j=1,2.)\label{metric}
\end{eqnarray}
with $x_{1}=M$, $x_{2}=P$. This line element measures the probability of a fluctuation between two states. Combining (\ref{horizon}) and (\ref{1entropy}), we obtain
a Bekenstein-Smarr-like formula
\begin{eqnarray}
 S=\pi(M+\sqrt{M^2-P^2})^2+4\pi P^{2}\ln (M+\sqrt{M^2-P^2}).
\end{eqnarray}
Thus, the Ruppeiner metric can be obtained through (\ref{metric}). {After a simple calculation}, the metric can be expressed, in terms of $r_+$ and $r_-$, as
\begin{eqnarray}
 g_{11}^{r}\!\!\!&=&\!\!\!\frac{8\pi\eta_{1} r_{+}}{(r_{+}-r_{-})^{3}},\nonumber\\
 g_{12}^{r}\!\!\!&=&\!\!\!g_{21}^{r}=-\frac{16\pi\sqrt{r_- r_+}(r_-^2+r_+r_-+r_+^2)}{(r_+-r_-)^3},\label{Rmetric}\\
 g_{22}^{r}\!\!\!&=&\!\!\!\frac{4\pi(6 r_-^3-5 r_+ r_-^2+10 r_+^2r_-+r_+^3)}{(r_+-r_-)^3}
            -8\pi\ln(r_+).\nonumber
\end{eqnarray}
It is clear that the metric is singular at $r_{+}=r_{-}$. So, the
Ruppeiner metric is useless to describe an extremal black hole and
we will restrict our discussion {for} the non-extremal black hole $r_{+}>r_{-}$. As shown in \cite{Ruppeiner00}, calculating the curvature may tell us the thermodynamic properties for a microscopic model. Although microscopic degree of freedom of a black hole is still unknown, we can perform with this technique, which may provide us with some useful information about the microstructure of the black hole (such as the correlation length). A direct calculation shows that the Ruppeiner curvature is
\begin{eqnarray}
 R_{r}=\frac{(2 r_-+r_+) (r_-^2+7 r_+ r_-+r_+^2)}{\pi\eta_{2}^2 r_{+}}.
\end{eqnarray}
{Here, the definitions of the Christoffel symbols and Riemann curvature tensor are the same as \cite{Ruppeiner1995rmp}, i.e., $\Gamma^{\lambda}_{\mu\nu}=\frac{1}{2}g^{\lambda\tau}(g_{\nu\tau,\mu}+g_{\mu\tau,\nu}-g_{\mu\nu,\tau})$ and $R^{\mu}_{\sigma\nu\tau}=\Gamma^{\mu}_{\sigma\nu,\tau}-\Gamma^{\mu}_{\sigma\tau,\nu}
                         +\Gamma^{\mu}_{\lambda,\tau}\Gamma^{\lambda}_{\sigma,\nu}
                         -\Gamma^{\mu}_{\lambda,\nu}\Gamma^{\lambda}_{\sigma,\tau}$.}
Note that $\eta_{2}$ contained a logarithmic term, which comes from the entropy. Thus, we can see that the logarithmic term has influence on the Ruppeiner curvature, as well as the thermodynamic metric (\ref{Rmetric}). We plot $R_{r}$ in Fig. \ref{curvatures} as functions of $r_{+}$ and $r_{-}$. Form it, we can see that the curvature $R_{r}$ is positive definite for all $r_{+}>r_{-}$ and it is divergent at $\eta_{2}=0$. The non-vanished curvature suggests that the statistical system of KS black hole is interacting. And the positive value implies that there is effective repulsive interaction among the microscopic particles carrying with the degrees of freedom. The divergent points also imply a phase transition, which is consistent with that of $C_{\Phi}$ and $C_{T}$. And the divergent points mean that unlimited repulsive force {appears}. This behavior is similar to the ideal gas of fermions, whose curvature is found to be positive and positive diverge at absolute zero \cite{Janyszek,Oshima}. The positive diverge points can be understood with the Pauli's exclusion principle, which forbids two particles in the same state with unlimited repulsive force. While for the ideal gas of bosons, the curvature is negative and {goes} to negative infinity at absolute zero \cite{Janyszek} appeared as Bose-Einstein condensation. On the other hand, the Ruppeiner curvature is regular at $\eta_{1}=0$ and $\eta_{3}=0$. So, it could not reflect the phase transition contained in $C_{P}$ and $C_{S}$. The reason of it is that the different choice of thermodynamic potential only gives us the thermodynamic stability and phase transitions in one thermodynamic ensemble. Thus, in order to obtain the thermodynamic stability and phase transitions in another ensemble, we should take another thermodynamic potential for the statistical system.

\section{Sumarry}

In this paper, we study the thermodynamic geometry and phase transition of KS black hole in the deformed HL gravity with coupling constant $\lambda=1$. We first calculate the four {type's} capacities. The heat capacity $C_{P}$ implies that the small KS black hole can stably exist in a heat bath, while the large one could not. $C_{P}$ is also found to be vanished at $r_+=r_-$, which corresponds to the extremal black hole, and $C_{P}$ is found to diverge at the points $\eta_{1}=0$ implying the existence of a phase transition. Other three capacities show the phase transition points at $\eta_{2}=0$, $\eta_{3}=0$, respectively. The points $\eta_{3}=0$ also show a more complicated structure than that of $\eta_{1}=0$, $\eta_{2}=0$. Then we examine the thermodynamic geometry. The results tell that the {phase transition points} are consistent with the singular points of the thermodynamic curvatures, i.e., the Ruppeiner curvature $R_{r}$ is singular at $\eta_{2}=0$. Therefore, we can conclude that the phase transition is included in the thermodynamic geometry. Moreover, the {thermodynamic curvature} $R_{r}$ produced from the Ruppeiner metric is positive definite for all $r_{+}>r_{-}$ and divergence at $\eta_{2}=0$. These suggest that the microstructure of the black hole has an effective repulsive interaction, which behaves like the ideal gas of fermions.

\acknowledgments
The authors would like to thank the anonymous referees whose comments largely helped us in improving the original manuscript. Wei would like to thank Dr. Kai-Nang Shao for his helpful discussions. This work was supported in part by the Huo Ying-Dong Education Foundation of Chinese Ministry of Education (No. 121106), the National Natural Science Foundation of China (No. 11075065), and the Fundamental Research Funds for the Central Universities (No. lzujbky-2012-k30).

\end{document}